\documentclass{icrc2009}

\usepackage{graphicx}   
\usepackage[font=footnotesize]{subfig} 
\usepackage{fixltx2e}
\usepackage{url}
\usepackage[usenames,dvipsnames]{color}
\usepackage{hyperref}
\definecolor{orange}{rgb}{1,0.5,0}
\hypersetup{
    colorlinks,%
    citecolor=RoyalBlue,%
    filecolor=RedViolet,%
    linkcolor=ProcessBlue,%
    urlcolor=RedViolet
}

\newcommand{\shorttitle}[1]%
{\markboth{Proceedings of the 31\MakeLowercase{$^{st}$} ICRC, {\L}\'{o}d\'{z} 2009}{#1} }
\newcommand{\etal}{\MakeLowercase{\textit{et al. }}} 

\begin{document}
\title{The Large Aperture GRB Observatory}

\author{ \vbox{ D.~Allard$^a$,
C.~Alvarez$^b$,
H.~Asorey$^c$,
H.~Barros$^d$,
\underline{X.~Bertou}$^{c,1}$,
M.~Castillo$^e$,
J.M.~Chirinos$^f$,
A.~De Castro$^d$,
S.~Flores$^g$,
J.~Gonzalez$^h$,
M.~Gomez Berisso$^c$,
J.~Grajales$^e$,
C.~Guada$^i$,
W.R.~Guevara Day$^j$,
J.~Ishitsuka$^g$,
J.A.~L\'opez$^k$,
O.~Mart\'inez$^e$,
A.~Melfo$^i$,
E.~Meza$^l$,
P.~Miranda Loza$^m$,
E.~Moreno Barbosa$^e$,
C.~Murrugarra$^d$,
L.A.~N\'u\~nez$^i$,
L.J.~Otiniano Ormachea$^j$,
G.~Perez$^n$,
Y.~Perez$^i$,
E.~Ponce$^e$,
J.~Quispe$^m$,
C.~Quintero$^i$,
H.~Rivera$^m$,
M.~Rosales$^i$,
A.C.~Rovero$^o$,
O.~Saavedra$^p$,
H.~Salazar$^e$,
J.C.~Tello$^d$,
R.~Ticona Peralda$^m$,
E.~Varela$^e$,
A.~Velarde$^m$,
L.~Villasenor$^q$,
D.~Wahl$^g$,
M.A.~Zamalloa$^r$
 (LAGO Collaboration)}\\
{$^a$}APC, CNRS et Universit\'e Paris 7. France\\
{$^b$}Universidad Autonoma de Chiapas, UNACH.  M\'exico \\
{$^c$}Centro At\'omico Bariloche, Instituto Balseiro.  Argentina \\
{$^d$}Laboratorio de F\'isica Nuclear, Universidad Sim\'on Bol\'ivar, Caracas. Venezuela \\
{$^e$}Facultad de Ciencias F\'isico-Matem\'aticas de la BUAP.  M\'exico \\
{$^f$}Michigan Technological University. USA\\
{$^g$}Instituto Geofisico del Per\'u, IGP. Lima - Per\'u\\
{$^h$}Universidad de Granada. Spain\\
{$^i$}Universidad de Los Andes, ULA. M\'erida - Venezuela\\
{$^j$}Comisi\'on Nacional de Investigaci\'on y Desarrollo Aeroespacial, CONIDA. San Isidro Lima - Per\'u\\
{$^k$}Universidad Central de Venezuela, Facultad de Ciencias, Departamento de F\'isica. Venezuela\\
{$^l$}Universidad Nacional de Ingenieria, UNI. Lima 25 - Per\'u\\
{$^m$}Instituto de Investigaciones F\'isicas, UMSA. Bolivia \\
{$^n$}Universidad Polit\'ecnica de Pachuca.  M\'exico \\
{$^o$}Instituto de Astronom\'ia y F\'isica del Espacio. Argentina \\
{$^p$}Dipartimento di Fisica Generale and INFN, Torino. Italy\\
{$^q$}Instituto de F\'isica y Matem\'aticas, Universidad de Michoac\'an. M\'exico\\
{$^r$}Universidad Nacional San Antonio Abad del Cusco. Per\'u
}

\shorttitle{X. Bertou \etal The Large Aperture GRB Observatory}

\maketitle

\begin{abstract}
The Large Aperture GRB Observatory (LAGO) is aiming at the detection of the
high energy (around 100\,GeV) component of Gamma Ray Bursts, using the single
particle technique in arrays of Water Cherenkov Detectors (WCD) in high
mountain sites (Chacaltaya, Bolivia, 5300\,m a.s.l., Pico Espejo, Venezuela,
4750\,m a.s.l., Sierra Negra, Mexico, 4650\,m a.s.l). WCD at high altitude offer
a unique possibility of detecting low gamma fluxes in the 10\,GeV - 1\,TeV range.
The status of the Observatory and data collected from 2007 to date will be
presented.
\footnotetext[1]{presenting and corresponding author, \href{mailto:bertou@cab.cnea.gov.ar}{bertou@cab.cnea.gov.ar}}

  \end{abstract}

\begin{IEEEkeywords}
Water Cherenkov Detector, Single Particle Technique, Gamma Ray Burst
\end{IEEEkeywords}

\section{Introduction}

Gamma Ray Burst are characterised by a sudden emission of electromagnetic
radiation at hard X-ray and soft $\gamma$-ray (X/$\gamma$) energies during a
short period of time, typically between 0.1 and 100 seconds. Since their discovery at the
end of the 60s by the VELA satellites\,\cite{kle}, GRB
have been of high interest to astrophysics.

They occur at an average rate of a few events per day, and their duration
shows a bimodal distribution with two different populations,
short duration GRBs (sGRB), characterised by durations of less than two seconds,
usually thought to be generated by the gravitational coalescence of two
compact objects (neutron stars or black holes) and
long duration GRBs (lGRB), usually associated with the
core  collapse ({\emph{collapsar}}) of a massive star, which tends to have a softer
spectrum than sGRB.

A first large data set of GRB was provided by the BATSE instrument on
board the Compton Gamma Rays Observatory (1991-2000).
BATSE GRBs incoming directions were isotropically
distributed with no evidence of clustering. The fluences observed were
furthermore incompatible with uniform distribution of sources, exhibiting a
deficit at low fluences.

GRBs origin was determined following afterglows
identification by Beppo-SAX (1996-2002). Due to better angular resolution than 
BATSE, afterglows could be detected at other wavelengths.
Spectroscopic measurements allowed the direct measurement of GRBs redshifts,
confirming they were cosmological in origin. 

Currently, GRB are registered by HETE, INTEGRAL, Swift and GLAST (renamed
Fermi Gamma-Ray Space Telescope). In the last 10 years, observation of the afterglows
allowed a much better understanding of the GRB phenomena. Most
observations have however been done below a few GeV of energy, and the
high energy (above 10 GeV) component in the GRB spectrum is
still poorly known.
Fermi/GLAST sensitivity has already provided some hints on the high energy 
component of GRBs\,\cite{GLASTHE}, and could allow to get individual GRB spectra up to 300
GeV should the flux of HE photon be above a few per m$^2$. In the meantime,
and at the highest energies where the flux is low, the only way to detect a
high energy emission of GRB is to work at ground level.

\section{GRB detection at ground level}
A classical method to use is called ``single particle technique'' (SPT)\,\cite{Aglietta}. When
high energy photons from a GRB reach the atmosphere, they produce
cosmic ray cascades. The energies are not enough
to produce a shower with many particles detectable at ground level (even at
high altitudes, only a few reach ground).  However, many photons are expected to arrive
during the burst, in a short period of time.
Should one have a ground array of particle detectors,
one would therefore see an
increase of the
background rate on all the detectors on this time scale.  This technique has already been unsuccessfully 
applied by INCA\,\cite{inca} in Bolivia and ARGO\,\cite{argo} in Tibet among others. A
general study of this technique can be found in \cite{ver}.
While affected by the atmospheric absorption (hence strongly dependant
on the zenith angle of observation), it is still the only available
method in the GeV energy range for ground based detectors.
Up to now, it has only been widely applied to arrays of scintillators or RPCs.
We have
already proposed using instead Water-Cherenkov Detectors\,\cite{auger}.
Their main advantage is their sensitivity to photons, which represent up
to 90\% of the secondary
particles at ground level for high energy photon initiated showers.

To get an idea of the potential of ground based observation, and the range of parameters
in which it can complement satellite observations, one can consider the
Chacaltaya Observatory, in Bolivia, at 5300\,m a.s.l., with a typical background rate of
secondaries of 3\,kHz/m$^2$ of WCD. A 5 $\sigma$ excess over background during 1 second would be a $5\times\sqrt{3000}\approx 270$ particle excess per m$^2$.
At this altitude, a 100\,GeV photon produces about 290 detectable particles in
a WCD\,\cite{alex}. Therefore, a fluence of one particle per m$^2$ at 100\,GeV could be seen
from the ground with a 1\,m$^2$ detector.

\begin{figure}[ht!]
\includegraphics[width=.5\textwidth]{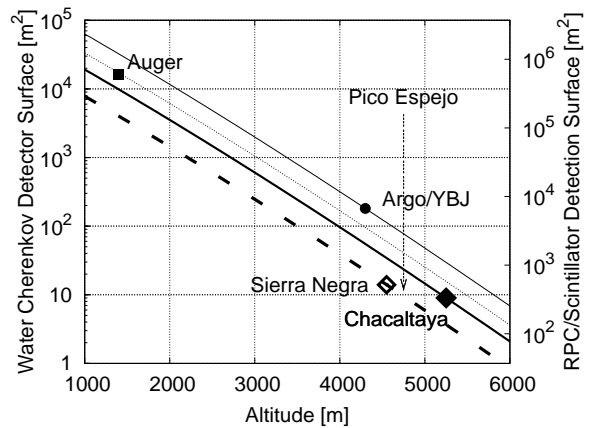}
\caption{
\label{fig:alt}
Lines of equal sensitivity for experiments of different size and altitude, neglecting geolatitude cutoff and assuming similar scaler threshold. Sierra Negra 
is shown as a dashed line, and Chacaltaya as a solid one. A few tens of m$^2$ of WCD at high altitude are as efficient as currently running experiments for the SPT.
}
\end{figure}

This SPT method has been tested on the largest WCD array
in operation, the Pierre Auger Observatory\,\cite{nimauger}. The sensitivity
of the Pierre Auger Observatory is however limited by its low altitude
(1400\,m a.s.l.) and should a burst be observed, the low bandwidth of each individual station would limit the scientific content of the results, as the only available data are integrated rates over one second. The
LAGO project compensates a much smaller area of detection by going for high
altitude sites, and uses a dedicated acquisition, optimised for
the SPT with rates being monitored on a short time scale. The three sites currently being instrumented are Sierra Negra (Mexico,
4550\,m\,a.s.l.), Chacaltaya (Bolivia, 5300\,m\,a.s.l.) and M\'erida
(Venezuela, 4765\,m\,a.s.l.).
It has previously been reported that
about 20\,m$^2$ of WCD in operation at Mount Chacaltaya 
would have the
same sensitivity as the full 16000\,m$^2$ of active surface of
Auger\,\cite{auger}. Figure \ref{fig:alt} shows the
equivalence between surface and altitude to get a similar sensitivity and
compares the LAGO sites with previous experiments.

\section{LAGO Experimental Setup}

Simulations were run to determine the optimal geometry of the WCDs.
The aperture gain for detectors of more than 4\,m$^2$ does not compensate
the increase in cost and difficulty to operate them, especially in remote
areas such as high altitude sites.
The chosen design is a $\approx$\,4\,m$^2$ WCD, with a central PMT, filled with water up to a level of 1.2\,m to 1.5\,m in order to
ensure a high probability of photon conversion in the water volume. The internal walls of the WCD
are covered by Tyvek\textsuperscript{\textregistered} or Banner-type material, to ensure a good
reflectivity and diffusivity. The PMT is connected to an acquisition board
from the prototype phase of the Pierre Auger
Observatory\,\cite{nimauger}. These boards provide 6 analog entries
which are sampled by 40\,MHz FADC allowing therefore up to 6 WCD to be
controlled by a single DAQ board. The digital signals are processed by
an APEX FPGA. An upgrade of these boards to 100 or 200\,MHz, for better
SPE counting, and an improved communication link is under way\,\cite{humb}.

The FPGA has been programmed to read out every 5\,ms the content of four scalers per
channel. The thresholds are set depending on the PMTs characteristics
(gain and noise). At Sierra Negra, they are set to about 15, 150 and 600 MeV
deposited in the WCD, while
a special scaler counts undershoots. At Chacaltaya and M\'erida, where higher gain
phototubes are available, they are set to $1/2$, 5 and 20 photoelectrons
(about 2, 25 and 100\,MeV deposited), with the same
undershoot counter. The undershoot counter is used to detect High Frequency
noise that could be picked-up by the PMTs cables during a lightning strike and
would be erroneously interpreted as many consecutive particles.

The data are then collected via a serial line by an acquisition PC, and
stored for data analysis. Replacing the acquisition PC by a single board PC
or a low cost laptop with SSD drive is foreseen to minimise the impact of the
harsh high altitude environment.

These data have a sampling rate of 5\,ms, much
smaller than what is usually used for the SPT. While this only marginally
lowers the detection threshold, it would provide crucial time structure
information should a burst be registered.

Currently, the Sierra Negra site is taking data since 2007
with three
4\,m$^2$ and two 1\,m$^2$ WCD.
PMT, DAQ PC failures and the rough hurricane season
limited the total useful  data accumulated since 2007 to the equivalent of 6 months of continuous data.
Two 4\,m$^2$ and one 1\,m$^2$ WCD have started operation in Chacaltaya in 2008 and are
in stable acquisition since beginning 2009.
A 3.5\,m$^2$ prototype and various smaller 1\,m$^2$ detectors are in operation at the Universidad
de los Andes, at 1600\,m\,a.s.l., and in Caracas (Venezuela). Installation at high
altitude is foreseen during 2009. More details on the LAGO sites and their
operation can be found in \cite{humb}.
A small 2\,m$^2$ prototype is instrumented at the Centro At\'omico Bariloche
(Argentina, 780\,m\,a.s.l.) and used for software development.
Two extra sites are under consideration in Peru, and a first prototype is
under construction in Lima. Colombia and the Himalaya are under study for new
installations.

\section{Data analysis and results}

\begin{figure}[ht!]
\includegraphics[width=.5\textwidth]{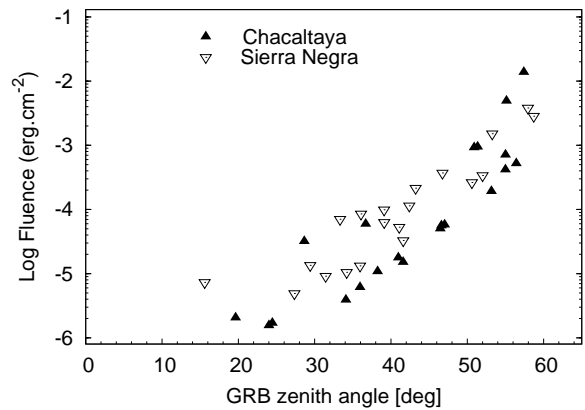}
\caption{
\label{fig:lim}
Fluence limits in the 0.5\,GeV - 100\,GeV range for the 41 bursts in the field of view of LAGO in the 2007
- April 2009 period, assuming a a spectral slope of -2.2. Filled up-triangles
are bursts occurring in Chacaltaya field of view. Empty down-triangles are bursts in Sierra Negra field of view.
}
\end{figure}

We searched for signal within 100 seconds of a GRB detected by satellites in
data taken from early 2007 to April 2009.
We
used the Gamma-Ray Burst Online Index (GRBOX\,\cite{grbox}) to extract bursts
data and selected those happening for each site with an apparent zenith angle
lower than 60 degrees. We requested a site to have at least 2 detectors in operation
at that moment, removing noisy detectors. This left us with 21 bursts for
Chacaltaya and 20 for Sierra Negra, with one burst occurring in the field of
view of both sites.
We then averaged the data in bins
of 100\,ms and looked for excesses (4\,$\sigma$ with $\sigma$ being the square
root of the average rate over 200 seconds before the burst) in coincidence in
at least 2 detectors. This left us with 2 bursts candidates for Chacaltaya
and 2 candidates for Sierra Negra. These were individually checked and found
consistent with statistical fluctuations. We then take the highest signal in
a 100\,ms bin to set a limit to the fluence between 0.5\,GeV and 100\,GeV
assuming a spectral slope of -2.2, based on simulations\cite{alex}. 
The fluence limits obtained are summarised in figure \ref{fig:lim}. The lowest
limit obtained is $1.6\times10^{-6}$\,erg.cm$^{-2}$ for GRB 080904.

Bursts can furthermore be searched independently of satellite data. However, should
such a burst be found it would be very difficult to attribute it to a cosmic
event and reject any possible instrument noise, unless a correlation is
found between sites. The current large angular separation between the two sites
of LAGO makes such a coincidence unlikely. New sites in between (Venezuela,
Peru, Colombia) will greatly increase this possibility.

\begin{figure}[ht!]
\includegraphics[width=.5\textwidth]{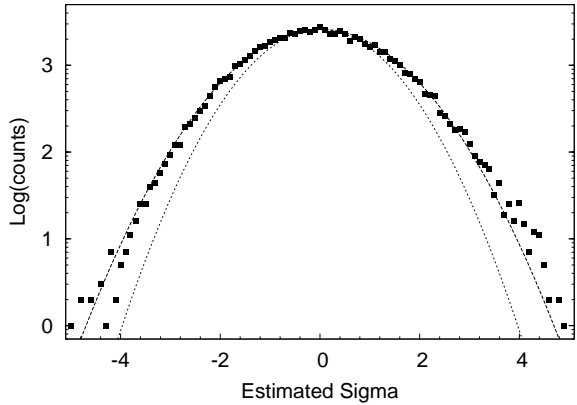}
\caption{
\label{fig:sigma}
Distribution of the estimated fluctuations by the sigma-delta method, with the
underlying Gaussian of $1.18\,\sigma$ width. A one $\sigma$ Gaussian is also
drawn for comparison.
}
\end{figure}

Nevertheless, an algorithm to search for potential bursts while rejecting
known noises has been developed and applied to the current data. Data are
averaged in 100\,ms bins and a running average is obtained by a sigma-delta
method, modifying the estimated average by 0.001\,Hz every time bin in the
direction of the rate of this bin. The second scaler of each channel is used
as the first one can be noisy on some detectors. The fluctuations
of each detector are assumed to be the square root of the estimated average.
The distribution of the fluctuation obtained by this method can be seen on
figure \ref{fig:sigma}. It is a Gaussian with width 1.18, due to correlated
noise and the method used to get the moving average.

A candidate burst is defined as an event where two detectors in coincidence see a 5 sigma
fluctuation (equivalent to 5.9 of our estimated fluctuation) at least twice in 
a 5 minute window. 16 candidate bursts are found in Chacaltaya, probably 
produced by electronic noise as signals are also found on a disconnected
channel (it is unlikely that these are true signals produced for example by
crosstalk as a GRB should manifest as many small signals and not by large PMT signals
which are the ones likely to produce electronic crosstalk). These candidate
bursts are likely HF noise produced by storms.

\begin{figure}[ht!]
\includegraphics[width=.5\textwidth]{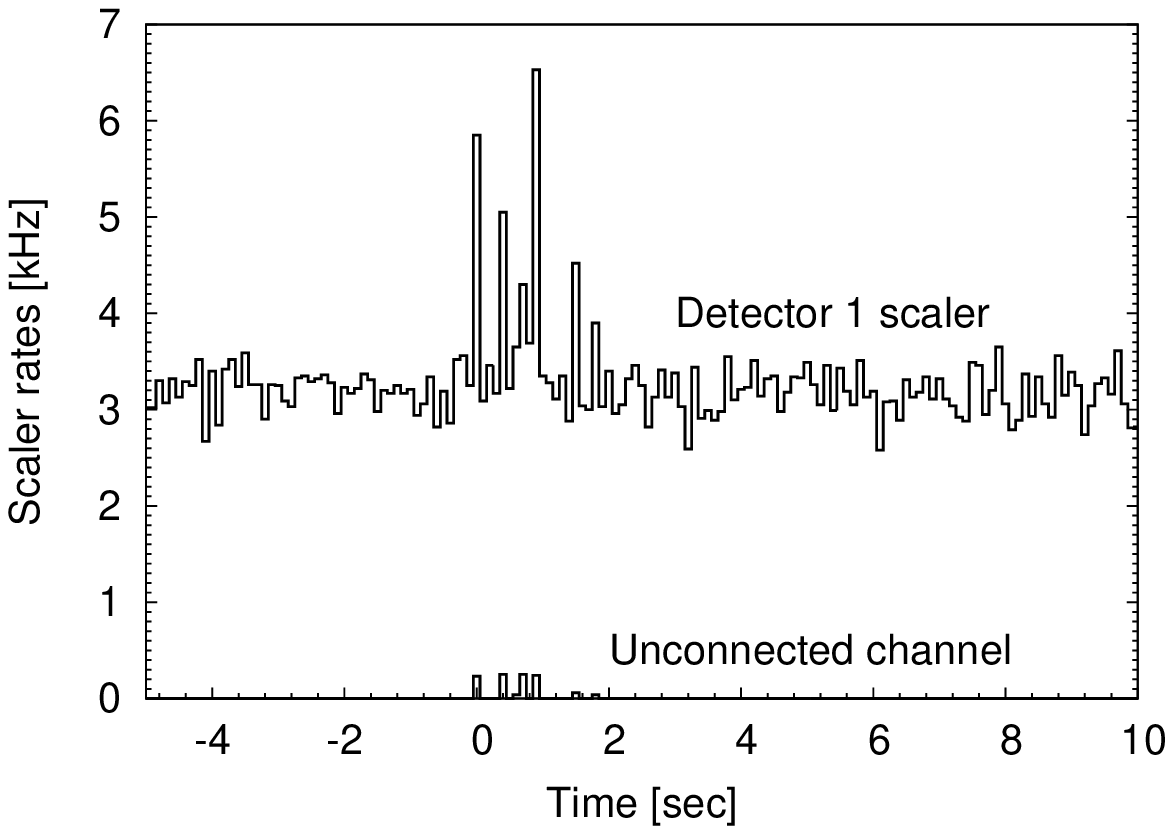}
\includegraphics[width=.5\textwidth]{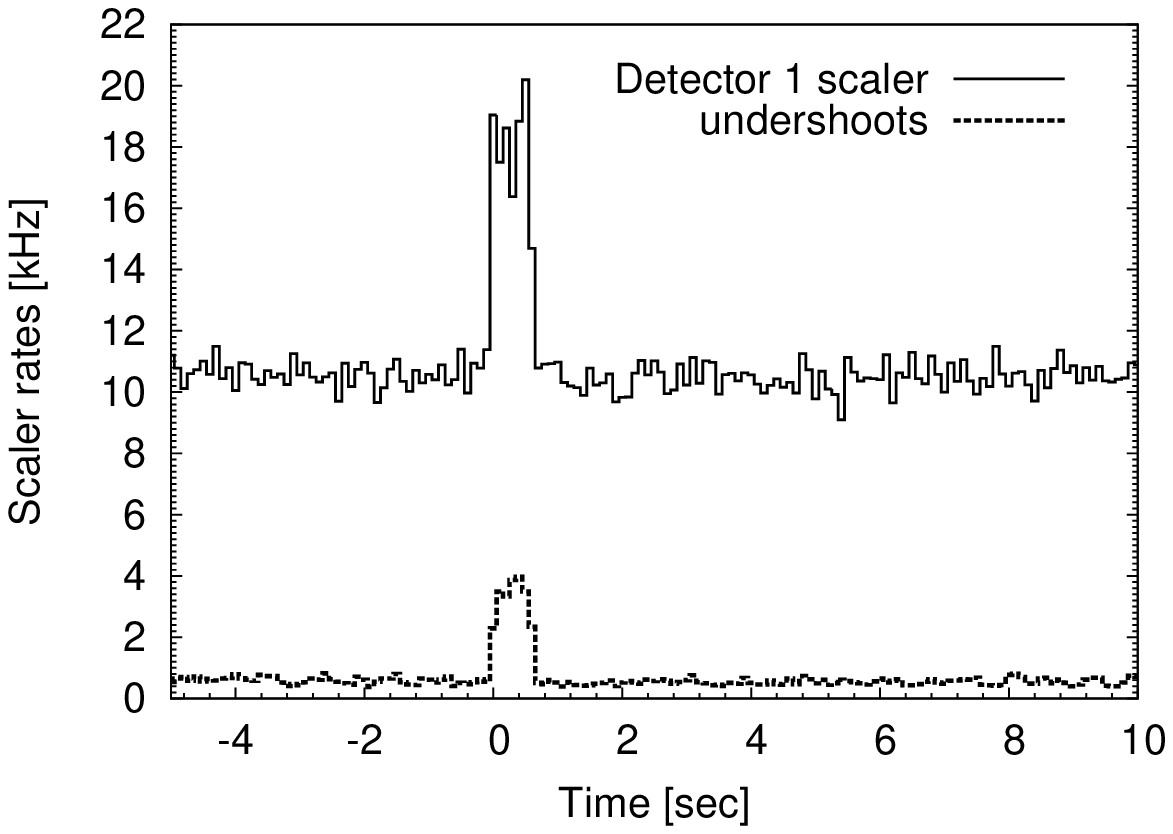}
\caption{
\label{fig:noise}
Example of noisy counting rates for Chacaltaya (top) and Sierra Negra (bottom).
One channel is shown, together with an unconnected channel (top) or the
undershoot counter (bottom). In both cases, signals in the other counter
indicate the burst is due to noise.
}
\end{figure}

The same analysis on the Sierra Negra data set provides a large set of 230
candidate bursts. While the Chacaltaya detectors
are installed inside a building under a thin roof, the Sierra Negra ones
are less protected and suffer more directly from the harsh weather conditions
of the site. Furthermore, Sierra Negra is quite isolated (together with the 
close-by Pico de Orizaba) in a vast plain while Chacaltaya is in a mountain
range. Finally, the hurricane season in Mexico is worse than the Bolivian
summer rains. All bursts are however rejected as HF noise candidates,
either using a disconnected channel or using the undershoot counter scaler.
Examples of these noisy events are given in figure \ref{fig:noise}.

\section{Conclusions}

The LAGO has been taking data since 2006 and is entering now in stable
data taking with two sites in operation.
The Sierra Negra site counts currently
with 14\,m$^2$ of calibrated and operating WCD, while 9\,m$^2$ of WCD are
taking data in Chacaltaya. Prototypes are in operation in 
M\'erida and Bariloche, while new ones are under construction in Peru.
Further sites are being investigated in Colombia, Guatemala and the Himalaya.

The data acquired since 2007 are of better quality than the one previously
reported\,\cite{prev} and a clean search for self-triggered bursts has been
done. No event out of HF noise has been found. 41 GRBs were reported by
satellites in the field of view of LAGO, and a specific search for excess
within 100 seconds of the burst was performed, with no excess found.
Limits were set for the fluences of these 41 bursts in the 0.5\,GeV - 100\,GeV range. The
lowest limit obtained is $1.6\times10^{-6}$\,erg.cm$^{-2}$,
comparable to what
the Pierre Auger Observatory can achieve\,\cite{resauger}.

In order to improve these limits, higher altitude sites are being looked for.
Higher gain PMTs, higher frequency sampling, and more
stable acquisition chain should also improve the data to be taken.

The LAGO project is very thankful to the Pierre Auger collaboration for
the lending of the engineering equipment.

\end{document}